\documentclass[journal=./achemso/jctc,manuscript=article,layout=onecolumn]{achemso}
\usepackage{comment} 
\usepackage{amsmath} 
\usepackage{amsfonts,amstext,amsgen,amsbsy,amsopn,amssymb} 

\usepackage{graphicx} 
\usepackage{subcaption} 
\usepackage{dcolumn} 
\usepackage{bm} 
\usepackage{physics}
\usepackage{makecell}
\usepackage{multirow}
\usepackage{tabstackengine}
\usepackage{qcircuit} 
\stackMath
\usepackage[dvipsnames, table]{xcolor}
	\definecolor{diagramColor}{RGB}{0,0,120}
\usepackage{tikz} 
	\usetikzlibrary{trees}
	\tikzset{
 		treenode/.style = {shape=rectangle, draw, diagramColor, ultra thick, align=center, 
                     top color=white, bottom color=white},
		root/.style     = {treenode, font=\normalsize},
  		env/.style      = {treenode, font=\normalsize},
		envtwo/.style      = {treenode, font=\normalsize},
		mode/.style = {edge from parent path={(\tikzparentnode.east) -- (\tikzchildnode.west)}}
	}
\usepackage{enumitem} 
\usepackage{mathtools} 

\usepackage[colorlinks=true,linkcolor=blue,citecolor=blue,urlcolor=black]{hyperref}
\usepackage{array}
\newcolumntype{?}{!{\vrule width 1.5pt}}
\usepackage{arydshln}

\newcommand{\halfcheckmark}[0]{\checkmark\raisebox{0.23em}{\kern-0.68em\large$\times$}}

\SectionNumbersOn

\title{Error-Mitigation Enabled Multicomponent Quantum Simulations Beyond the Born–Oppenheimer Approximation}

\author{Delmar G. A. Cabral}
\affiliation{Department of Chemistry, Yale University, New Haven, CT 06520, U.S.A.}
\author{Brandon Allen}
\affiliation{Department of Chemistry, Yale University, New Haven, CT 06520, U.S.A.}
\author{Fabijan Pavošević}
\affiliation{Algorithmiq Ltd., Kanavakatu 3C, FI-00160 Helsinki, Finland}
\author{Sharon Hammes-Schiffer}
\affiliation{Department of Chemistry, Princeton University, Princeton, NJ  08540, U.S.A.}
\author{Pablo Díez-Valle}
\affiliation{Instituto Tecnológico de Galicia, Cantón Grande 9, Planta 3, 15003 A Coruña, Spain}
\author{Jack S. Baker}
\affiliation{LG Electronics Toronto AI Lab, Toronto, Ontario M5V 1M3, Canada}
\author{Gaurav Saxena}
\affiliation{LG Electronics Toronto AI Lab, Toronto, Ontario M5V 1M3, Canada}
\author{Thi Ha Kyaw}
\affiliation{LG Electronics Toronto AI Lab, Toronto, Ontario M5V 1M3, Canada}
\email{thiha.kyaw@lge.com}
\author{Victor S. Batista}
\affiliation{Department of Chemistry, Yale University, New Haven, CT 06520, U.S.A.}
\alsoaffiliation{Yale Quantum Institute, Yale University, New Haven, CT 06511, U.S.A.} 
\email{victor.batista@yale.edu}

\date{\today}

\begin{document}

\maketitle

\begin{abstract}
We introduce a multicomponent unitary coupled cluster framework for quantum simulations of molecular systems that incorporate both electronic and nuclear quantum effects beyond the Born–Oppenheimer approximation. Using the nuclear–electronic orbital formalism, we construct mcUCC ansätze for positronium hydride and molecular hydrogen with a quantum proton, and analyze hardware requirements for different excitation truncations. To further reduce resource costs effectively, we employ the local unitary cluster Jastrow ansatz and implement it experimentally on IBM Q’s Heron superconducting hardware. 
With the Physics-Inspired Extrapolation error mitigation protocol, the computed ground-state energies remain within chemical accuracy, consistent with the stated uncertainty level.
These results provide the first demonstration of error-mitigated multicomponent correlated simulations on quantum hardware and outline a path toward scalable algorithms unifying electronic and nuclear degrees of freedom. 
\end{abstract}

\section{Introduction}

Many fundamental chemical processes involve nuclear quantum effects that challenge the Born--Oppenheimer (BO) separation of electronic and nuclear motion. Phenomena such as proton tunneling, hydrogen transfer, and proton-coupled electron transfer are governed by zero-point motion and nonadiabatic couplings that significantly influence reaction thermodynamics and kinetics~\cite{Waluk2024, Meisner2016, HammesSchiffer2015, Tyburski2021, Klinman2013, HammesSchiffer2023}. Classical treatments that confine nuclei to point particles can therefore produce qualitatively incorrect predictions.  

The nuclear--electronic orbital (NEO) framework addresses this limitation by treating selected light nuclei (typically protons) quantum mechanically alongside electrons. By incorporating nuclear delocalization, tunneling, and electron--nucleus correlation directly into the wavefunction, NEO theory provides a unified and systematically improvable approach for systems in which electronic and nuclear degrees of freedom are strongly coupled~\cite{webb2002multiconfigurational,HammesSchiffer_2021}.  

Despite the accuracy of NEO-based methods, solving correlated electronic--nuclear structure problems remains computationally demanding. Classical algorithms for exact solutions scale exponentially with system size, motivating the use of quantum computation as an alternative route to scalable molecular simulation~\cite{abrams_simulation_1997_hubbard, aspuru-guzik_simulated_2005}. Two major paradigms have emerged: quantum phase estimation (QPE) and the variational quantum eigensolver (VQE)~\cite{bauer_quantum_2020, cao_quantum_2019, Bharti2022Feb}. QPE can, in principle, yield exact eigenvalues efficiently, but its deep, coherent circuits remain impractical for current noisy intermediate-scale quantum (NISQ) devices~\cite{Preskill2018Aug}. VQE, in contrast, employs shallow parameterized circuits with classical optimization and has been successfully demonstrated on small molecular systems~\cite{peruzzo_variational_2014, omalley_scalable_2016, Kandala2017Sep, romero_strategies_2019, hempel_quantum_2018}.  

Within VQE, the unitary coupled cluster (UCC) ansatz~\cite{Anand2022Mar}, derived from the coupled cluster formalism~\cite{Coester1960Jun}, provides a physically grounded and systematically extensible representation of correlated wavefunctions. While classical simulations of UCC scale exponentially, its quantum implementation requires only polynomial resources, making it a natural foundation for chemically realistic quantum algorithms.  

Recent work has unified NEO theory with quantum algorithms via multicomponent UCC (mcUCC) ans\"atze, enabling beyond-BO simulations that explicitly include quantum nuclear motion in both the reference and variational spaces~\cite{mcUCC_NEO, Kovyrshin2023, nykanen2023toward, kovyrshin2025approximatequantumcircuitcompilation, Kovyrshin2023_long, Culpitt2025}. This NEO--quantum computing (NEO-QC) framework provides a rigorous route to first-principles modeling of electronic--nuclear correlated systems, including positronic and protonic species, where classical approximations fail.  

However, the accuracy of hybrid quantum--classical algorithms on present-day devices remains constrained by noise and decoherence. Full quantum error correction is not yet feasible, but quantum error mitigation (QEM) methods can substantially reduce systematic bias without fault-tolerant overhead. Among the most widely used are zero-noise extrapolation (ZNE), which estimates noise-free observables by deliberate noise amplification and extrapolation~\cite{Li_2017, Temme_2017}, and probabilistic error cancellation (PEC), which reconstructs unbiased estimators via quasi-probability sampling~\cite{Temme_2017}. Complementary strategies such as symmetry verification~\cite{BonetMonroig_2018} and virtual (state) distillation~\cite{Huggins_2021} have further improved accuracy in quantum chemistry experiments on real hardware~\cite{Arute_2020, McCaskey_2019, Lolur_2023}.  

In this work, we employ a recently developed, physically motivated error mitigation approach, Physics-Inspired Extrapolation (PIE), which extends the ZNE framework by deriving its functional form from restricted quantum dynamics~\cite{DiezValle_2025, Saxena2024Sep}. PIE provides an interpretable extrapolation model, mitigates overfitting, and reduces sampling overhead relative to polynomial ZNE, enabling chemically accurate energy estimates for beyond-BO benchmarks on current NISQ hardware.  

The remainder of this paper is organized as follows. Section~\ref{sec:NEO_framework} introduces the NEO Hamiltonian and working equations. Section~\ref{sec:coupled_cluster_formalism} describes the mcUCC ansatz used in classical VQE simulations, which serve as surrogates for near-term quantum implementations limited by circuit depth and hardware noise. Section~\ref{sec:VQE} outlines the VQE framework and Hamiltonian constructions for PsH and HHq systems. Section~\ref{sec:results_and_discussion} presents PIE-based, error-mitigated VQE results using the LUCJ ansatz on IBM Q’s Heron device. Finally, Section~\ref{sec:concluding_remarks} summarizes the results and discusses prospects for scalable, multicomponent quantum simulations beyond the BO approximation.

\section{Nuclear Electronic Orbital (NEO) Framework \label{sec:NEO_framework}}

Building upon prior work establishing the foundations of mcUCC methods for quantum computation~\cite{mcUCC_NEO}, we perform simulations of the same systems on a quantum simulator and analyze the computational requirements for execution on real NISQ devices. Two model systems are considered: molecular hydrogen with one quantum mechanical proton ($\mathrm{HHq}$) and positronium hydride ($\mathrm{PsH}$) (see Fig.~\ref{fig:H2_PsH_scheme}).

In the NEO formalism, both electrons and selected light nuclei (e.g., protons or positrons) are treated quantum mechanically, while heavier nuclei remain classical. The total NEO Hartree--Fock (NEO-HF) wavefunction is expressed as a product of electronic and nuclear components,
\begin{equation}
    |\Psi_{\text{NEO-HF}} (\chi_e, \chi_p) \rangle 
    = |\Phi_e(\chi_e)\rangle \otimes |\Phi_p(\chi_p)\rangle,
\end{equation}
where $\Phi_e(\chi_e)$ and $\Phi_p(\chi_p)$ are the electronic and quantum-nuclear wavefunctions, each expanded in their respective molecular orbital bases, $\chi_e$ and $\chi_p$.

The total Hamiltonian for a system containing electrons, quantum nuclei, and classical nuclei is written as
\begin{align}\label{neo_hamiltonian}
    \hat{H}_{\text{NEO}} 
    &= \hat{T}_e + \hat{T}_p + \hat{V}_{eN} + \hat{V}_{pN}
     + \hat{V}_{ee} + \hat{V}_{pp} + \hat{V}_{ep} + V_{NN},
\end{align}
where $\hat{T}_e$ and $\hat{T}_p$ are the kinetic energy operators for electrons and quantum nuclei, respectively, and the $\hat{V}$ terms represent the corresponding Coulomb interactions.  

The one-particle operators are defined as
\begin{align}
    \hat{T}_e + \hat{V}_{eN} &=
        \sum_{i=1}^{N_e}
        \left(
            -\frac{1}{2m_e}\nabla_i^2
            - \sum_{a=1}^{N_{\text{nuc}}} \frac{Z_a}{|\mathbf{R}_a - \mathbf{r}_i^{e}|}
        \right), \\
    \hat{T}_p + \hat{V}_{pN} &=
        \sum_{I=1}^{N_p}
        \left(
            -\frac{1}{2m_p}\nabla_I^2
            + \sum_{a=1}^{N_{\text{nuc}}} \frac{Z_a}{|\mathbf{R}_a - \mathbf{r}_I^{p}|}
        \right),
\end{align}
while the two-particle Coulomb interactions take the form
\begin{align}
    \hat{V}_{ee} &= 
        \sum_{i<j}^{N_e} \frac{1}{|\mathbf{r}_i^{e} - \mathbf{r}_j^{e}|}, \\
    \hat{V}_{pp} &= 
        \sum_{I<J}^{N_p} \frac{1}{|\mathbf{r}_I^{p} - \mathbf{r}_J^{p}|}, \\  
    \hat{V}_{ep} &= 
        -\sum_{i=1}^{N_e} \sum_{I=1}^{N_p}
        \frac{1}{|\mathbf{r}_i^{e} - \mathbf{r}_I^{p}|}.
\end{align}
Here, indices $i,j$ denote electrons, $I,J$ denote quantum nuclei, and $\mathbf{r}$ represents particle coordinates. $Z_a$ and $\mathbf{R}_a$ denote the charge and position of the $a^{\text{th}}$ classical nucleus, $\nabla$ is the Laplacian operator, and $m_e$ and $m_p$ are the electron and quantum-nuclear masses, respectively.
The classical nucleus–nucleus repulsion is given by
\begin{align}
    V_{NN} = 
        \sum_{a<b}^{N_{\text{nuc}}}
        \frac{Z_a Z_b}{|\mathbf{R}_a - \mathbf{R}_b|},
\end{align}
which is constant for a fixed nuclear geometry.

For practical computations, these operators are expressed in a finite one-particle basis $\{\phi\}$ in terms of one- and two-particle integrals. The one-particle matrix elements are
\begin{align}
    h_{ij} &=
        \int \phi_i^*(\mathbf{r}^{e})
        \left(
            -\frac{1}{2m_e}\nabla^2
            - \sum_{a=1}^{N_{\text{nuc}}} \frac{Z_a}{|\mathbf{R}_a - \mathbf{r}^{e}|}
        \right)
        \phi_j(\mathbf{r}^{e})\, d\mathbf{r}^{e}, \\
    h_{IJ} &=
        \int \phi_I^*(\mathbf{r}^{p})
        \left(
            -\frac{1}{2m_p}\nabla^2
            + \sum_{a=1}^{N_{\text{nuc}}} \frac{Z_a}{|\mathbf{R}_a - \mathbf{r}^{p}|}
        \right)
        \phi_J(\mathbf{r}^{p})\, d\mathbf{r}^{p},
\end{align}
and the two-particle integrals are
\begin{align}
    h_{ijkl} &=
        \int \phi_i^*(\mathbf{r}_1^{e}) \phi_j^*(\mathbf{r}_2^{e})
        \frac{1}{|\mathbf{r}_1^{e} - \mathbf{r}_2^{e}|}
        \phi_k(\mathbf{r}_1^{e}) \phi_l(\mathbf{r}_2^{e})\,
        d\mathbf{r}_1^{e} d\mathbf{r}_2^{e}, \\
    h_{IJKL} &=
        \int \phi_I^*(\mathbf{r}_1^{p}) \phi_J^*(\mathbf{r}_2^{p})
        \frac{1}{|\mathbf{r}_1^{p} - \mathbf{r}_2^{p}|}
        \phi_K(\mathbf{r}_1^{p}) \phi_L(\mathbf{r}_2^{p})\,
        d\mathbf{r}_1^{p} d\mathbf{r}_2^{p}, \\
    h_{iIkJ} &=
        \int \phi_i^*(\mathbf{r}^{e}) \phi_I^*(\mathbf{r}^{p})
        \frac{1}{|\mathbf{r}^{e} - \mathbf{r}^{p}|}
        \phi_k(\mathbf{r}^{e}) \phi_J(\mathbf{r}^{p})\,
        d\mathbf{r}^{e} d\mathbf{r}^{p}.
\end{align}

This multicomponent Hamiltonian mirrors the structure of the conventional electronic Hamiltonian, with distinct one- and two-particle contributions, while extending it to include explicit electronic–nuclear and nuclear–nuclear interaction terms. Each operator term can be evaluated using standard electronic structure integrals and serves as a foundation for correlated methods that improve upon the mean-field NEO-HF energy bound.

\section{Coupled Cluster Formalism \label{sec:coupled_cluster_formalism}}

Electronic--nuclear correlation beyond the NEO-HF level can be systematically included using the coupled cluster (CC) approach, which provides a size-extensive and systematically improvable description of correlated wavefunctions. The NEO-HF wave function serves as the reference state for constructing correlated multicomponent wave functions.

Gate-based quantum computers naturally implement unitary operations, making the UCC ansatz an especially suitable form of the CC method for quantum algorithms. The NEO-UCC wavefunction is generated by applying a unitary exponential operator~\cite{bartlett_alternative_1989} to the NEO-HF reference:
\begin{equation}
    |\Psi_{\mathrm{NEO\text{-}UCC}}\rangle 
    = e^{\hat{T} - \hat{T}^\dagger} 
    |\Psi_{\mathrm{NEO\text{-}HF}}\rangle,
\end{equation}
where $\hat{T}$ is the excitation operator, composed of fermionic creation ($a^\dagger$) and annihilation ($a$) operators weighted by variational amplitudes $t$. That is,
\begin{align}
    \hat{T} &= \hat{T}_1 + \hat{T}_2 + \hat{T}_3 + \cdots , \label{coupled_cluster_expansion}
\end{align}
where the subscript on each $T$ denotes the order of the excitation operator. For example: $T_1 = \sum_{ia}t_i^aa_a^\dagger a_i + \sum_{IA}t_I^Aa_A^\dagger a_I$, $T_2 = 1/4\sum_{ijab}t_{ij}^{ab}a_a^\dagger a_b^\dagger a_j a_i + 1/4\sum_{IJAB}t_{IJ}^{AB}a_A^\dagger a_B^\dagger a_J a_I + \sum_{iI}^{aA} t_{iI}^{aA} a_a^\dagger a_A^\dagger a_I a_i$ and so on. Here, indices $i,j,\ldots$ refer to occupied electronic orbitals, and  $a,b,\ldots$ to virtual electronic orbitals. For quantum nuclei (e.g., protons) or positrons, uppercase indices $I,J,\ldots$ and $A,B,\ldots$ denote occupied and virtual protonic/positronic orbitals, respectively.

In practice, the cluster operator in Eq.~\ref{coupled_cluster_expansion} is truncated to singles and doubles (UCCSD) due to the exponential scaling of higher-order excitations. The corresponding NEO-UCC energy is obtained by minimizing the expectation value of the NEO Hamiltonian \cite{Pavosevic2022Jun,mcUCC_NEO} with respect to all variational amplitudes:
\begin{equation}
    E_{\mathrm{NEO\text{-}UCC}} 
    = \underset{\{t\}}{\mathrm{min}}
      \langle \Psi_{\mathrm{NEO\text{-}UCC}} |
      \hat{H}_{\mathrm{NEO}} |
      \Psi_{\mathrm{NEO\text{-}UCC}} \rangle.
\end{equation}
This variational formulation allows straightforward integration into hybrid quantum--classical algorithms such as the VQE, where the amplitudes $\{t\}$ are optimized iteratively using energy feedback from a quantum device.

\section{Variational Quantum Eigensolver Method \label{sec:VQE}}

\subsection{Variational Quantum Eigensolver in the NEO framework}

\begin{figure}[ht!]
    \centering
    \includegraphics[width=\linewidth]{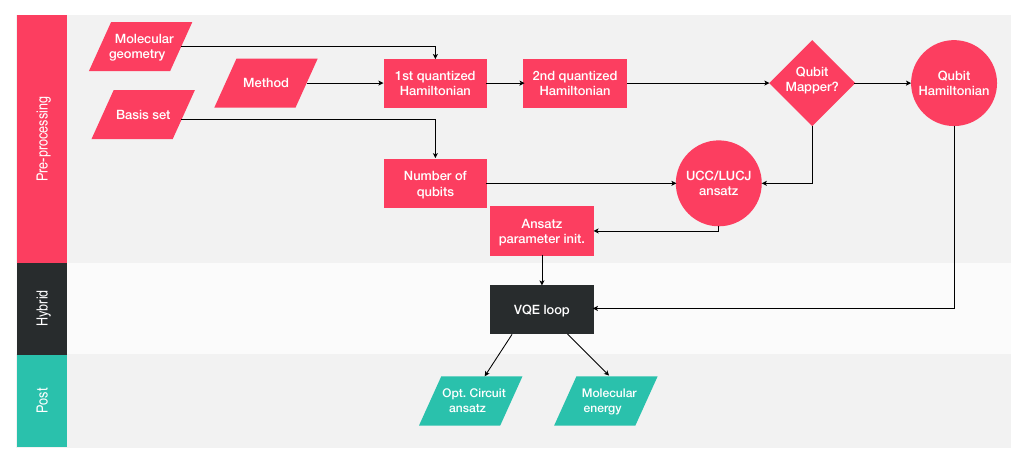}
    \caption{Schematic representation of the (VQE) algorithm. The molecular Hamiltonian is precomputed classically, while the quantum device evaluates expectation values and supplies energy feedback to the classical optimizer.}
    \label{fig:VQE_algorithm}
\end{figure}

The inclusion of electronic--nuclear correlation effects within a quantum computing framework can be achieved using the hybrid quantum--classical VQE algorithm~\cite{peruzzo_variational_2014, omalley_scalable_2016, romero_strategies_2019, hempel_quantum_2018}, schematically illustrated in Fig.~\ref{fig:VQE_algorithm}. In this approach, the molecular Hamiltonian is precomputed classically and encoded in a qubit representation, while a quantum processor evaluates expectation values for a parametrized trial wavefunction. The variational parameters are iteratively optimized by a classical optimizer to minimize the total energy until self-consistency is achieved.

In this work, the NEO-HF problem is first solved classically to obtain the reference orbitals and molecular integrals for a given molecular geometry and basis set. Each operator term in the NEO Hamiltonian (Eq.~\ref{neo_hamiltonian}) is evaluated in the atomic orbital basis and subsequently transformed into the molecular orbital basis through a standard congruence transformation. The resulting Hamiltonian is then expressed in its second-quantized form.

The second-quantized Hamiltonian is mapped to qubit operators using a fermion-to-qubit transformation. Common mappings include the Jordan--Wigner~\cite{JordanWigner1928, JordanWigner_application_PhysRevA.65.042323} and Bravyi--Kitaev~\cite{Bravyi_Kitaev_doi:10.1063/1.4768229} transformations. In this work, both mappings were implemented using the \texttt{OpenFermion} library~\cite{openfermion} (\verb|openfermion.transforms.jordan_wigner| and \verb|openfermion.transforms.bravyi_kitaev|), with results reported primarily using the Jordan--Wigner transformation.

Parallel to the Hamiltonian encoding, the multicomponent CC excitation operators (singles, doubles, and selected triples) are constructed classically following the mcUCC formalism~\cite{mcUCC_NEO}. These operators define the parameterized trial state used in VQE, typically in the form of a UCC ansatz with variational amplitudes serving as tunable parameters.

During VQE execution, the quantum circuit corresponding to the chosen ansatz is evaluated on a quantum backend or simulator, such as those provided in \texttt{Qiskit}~\cite{Qiskit}. The expectation value of the energy is computed as
\begin{equation}
    E(\boldsymbol{\theta}) = 
    \langle \Psi(\boldsymbol{\theta}) | \hat{H}_{\mathrm{NEO}} | \Psi(\boldsymbol{\theta}) \rangle,
\end{equation}
where $\boldsymbol{\theta}$ represents the set of variational parameters. A classical optimizer (e.g., COBYLA or SPSA) updates $\boldsymbol{\theta}$ iteratively to minimize the energy:
\begin{equation}
    E_{\text{min}} = \underset{\boldsymbol{\theta}}{\mathrm{min}}\, E(\boldsymbol{\theta}).
\end{equation}
This iterative feedback loop continues until the convergence criterion is met, yielding the lowest energy consistent with the chosen ansatz and hardware precision.

\subsection{Implementation for Representative Systems}

\begin{figure}[ht!]
    \centering
    \includegraphics[width=0.7\linewidth]{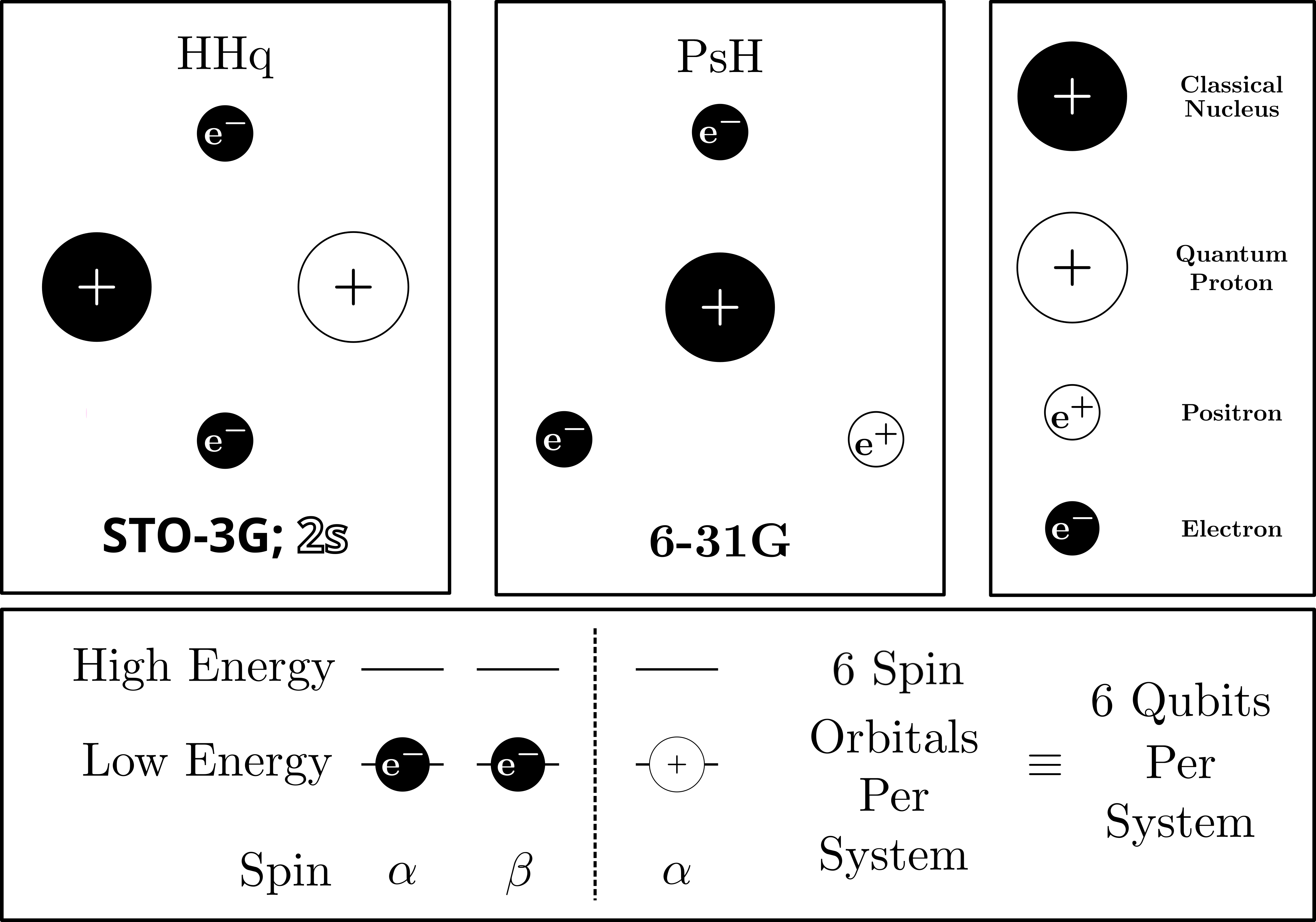}
    \caption{Top: Schematic representations of the hydrogen molecule with a quantum mechanical proton ($\mathrm{HHq}$) and positronium hydride ($\mathrm{PsH}$).  
    Bottom: Spin–orbital configurations for both systems under the chosen bases: 6-31G for electronic and positronic orbitals in $\mathrm{PsH}$; STO-3G for electronic orbitals and 2s for protonic orbital in $\mathrm{HHq}$.}
    \label{fig:H2_PsH_scheme}
\end{figure}

To enable direct comparison with prior multicomponent studies, our simulations are performed within a minimal basis framework. Specifically, the 6-31G basis set\cite{6-31G} is used for all quantum particles in positronium hydride (PsH), while for dihydrogen ($\mathrm{HHq}$), the electronic orbitals are described using the STO-3G basis set\cite{sto3g} and the quantum proton with a dedicated 2s basis set\cite{mcUCC_NEO} (Fig.~\ref{fig:H2_PsH_scheme}). Although this work focuses on these minimal systems, the same workflow is readily extendable to larger basis sets and other multicomponent molecular systems.

Each system comprises six spin–orbitals and three quantum particles:  
(i) $\mathrm{PsH}$, containing two quantum electrons and one quantum positron; and  
(ii) $\mathrm{HHq}$, containing two quantum electrons and one quantum nucleus.  
Each electron contributes two spin–orbitals, while the positron or quantum proton contributes two spatial orbitals. The orbital arrangement and particle composition are depicted in Fig.~\ref{fig:H2_PsH_scheme}.

To reduce circuit depth and gate count for near-term quantum implementation, we consider two strategies.  
First, only spin-conserving excitations are included for the electronic subspace, while the positron and quantum proton are both fixed to the alpha spin state. 
In a minimal basis, the second-quantized CC excitation operators restricted to spin-allowed transitions are given by:
\begin{align}
    t_{1e} &= (t_{0}^{2} c^\dagger_2 c_0 - t_{2}^{0} c^\dagger_0 c_2)
    + (t_{1}^{3} c^\dagger_3 c_1 - t_{3}^{1} c^\dagger_1 c_3),  \label{singles_e} \\
    t_{1p} &= (t_{4}^{5} c^\dagger_5 c_4 - t_{5}^{4} c^\dagger_4 c_5), \label{singles_p} \\
    t_{2ee} &= (t_{01}^{23} c^\dagger_2 c^\dagger_3 c_1 c_0 - t_{23}^{01} c^\dagger_0 c^\dagger_1 c_3 c_2), \label{doubles_ee} \\
    t_{2ep} &= (t_{04}^{25} c^\dagger_2 c^\dagger_5 c_4 c_0 - t_{25}^{04} c^\dagger_0 c^\dagger_4 c_5 c_2)
    + (t_{14}^{35} c^\dagger_3 c^\dagger_5 c_4 c_1 - t_{35}^{14} c^\dagger_1 c^\dagger_4 c_5 c_3), \label{doubles_ep} \\
    t_{3eep} &= (t_{014}^{235} c^\dagger_2 c^\dagger_3 c^\dagger_5 c_4 c_1 c_0 - t_{235}^{014} c^\dagger_0 c^\dagger_1 c^\dagger_4 c_5 c_3 c_2). \label{triples_eep}
\end{align}
Here, indices $0$–$3$ label electronic spin orbitals, while $4$ and $5$ correspond to the protonic (or positronic) spatial orbitals.

As an alternative strategy to operator selection, we also employ the ADAPT-VQE framework~\cite{AdaptVQE2019_original}.  
Unlike conventional VQE, ADAPT-VQE dynamically constructs the ansatz by iteratively selecting the operator from a predefined pool that yields the largest energy gradient at each step.  
This adaptive procedure typically results in a more compact ansatz, often with significantly fewer operators than the full operator pool, thereby reducing both circuit depth and total quantum resource requirements.

\section{Results and Discussion \label{sec:results_and_discussion}}

\subsection{Classical Simulations}

Classical simulations were performed to evaluate different combinations of cluster excitation operators, categorized by particle type (electronic or protonic/positronic) and excitation order (singles, doubles, triples). The simulations were executed on the \texttt{FakeNairobiV2} backend, a seven-qubit architecture containing one additional qubit beyond the minimum required for these systems. This device was used to assess circuit composition and resource requirements prior to deployment on real quantum hardware. Table~\ref{tab:hhq_PsH_JW} summarizes the gate counts, circuit depths, and corresponding VQE energies for each operator pool. The benchmark energy was obtained from a NEO full configuration interaction (NEO-FCI) calculation.

It is evident that the highest accuracy is obtained when all excitation operators are included; however, the resulting circuit depth and gate count far exceed the practical limits of current NISQ devices. Reducing the operator pool by truncating higher-order or mixed excitations decreases computational cost but also limits the recoverable correlation energy. When only single electronic and protonic excitations ($t_{1e}, t_{1p}$) are included, the computed energy is identical to the Hartree--Fock reference, confirming that individual single excitations do not contribute to correlation energy in these systems.  

\begin{table}[ht!]
    \centering
    \caption{Energies, circuit depths, and gate counts for the VQE simulations of $\mathrm{HHq}$ and $\mathrm{PsH}$ systems using the Jordan--Wigner qubit mapping.}
    \label{tab:hhq_PsH_JW}
    \begin{tabular}{lccccccc}
        \hline
        HHq Operator Pool & RZ & SX & CNOT & X & Total & Depth & Energy (Hartree) \\
        \hline
        $t_{1e}$,$t_{1p}$ & 55 & 48 & 47 & 1 & 151 & 112 & -1.059569 \\
        $t_{1p}$,$t_{2ee}$ & 109 & 86 & 68 & 2 & 265 & 178 & -1.079396 \\
        $t_{1e}$,$t_{2ee}$ & 144 & 113 & 80 & 3 & 340 & 225 & -1.079406 \\
        $t_{2ee}$,$t_{2ep}$ & 211 & 170 & 115 & 4 & 500 & 329 & -1.079421 \\
        $t_{1e}$,$t_{1p}$,$t_{2ee}$,$t_{2ep}$ & 246 & 192 & 129 & 6 & 573 & 379 & -1.079431 \\
        $t_{1e}$,$t_{1p}$,$t_{2ee}$,$t_{2ep}$,$t_{3eep}$ & 499 & 380 & 227 & 12 & 1118 & 743 & -1.079433 \\
        \hline
        LUCJ ansatz (numerical) & 39 & 20 & 16 & 8 & 83 & 25 & -1.079406 \\
        LUCJ ansatz (experimental) & & & & & & & $-1.077 \pm 0.009$ \\
        \hline
        NEO-HF (classical) & & & & & & & -1.059569 \\
        NEO-FCI (classical) & & & & & & & -1.079434 \\
        \hline\hline
        PsH Operator Pool & RZ & SX & CNOT & X & Total & Depth & Energy (Hartree) \\
        \hline
        $t_{1e}$,$t_{1p}$ & 55 & 48 & 47 & 1 & 151 & 112 & -0.558727 \\
        $t_{1p}$,$t_{2ee}$ & 107 & 86 & 68 & 2 & 263 & 175 & -0.569124 \\
        $t_{1e}$,$t_{2ee}$ & 144 & 113 & 80 & 3 & 340 & 224 & -0.569124 \\
        $t_{2ee}$,$t_{2ep}$ & 202 & 166 & 115 & 5 & 488 & 328 & -0.572710 \\
        $t_{1e}$,$t_{1p}$,$t_{2ee}$,$t_{2ep}$ & 234 & 188 & 129 & 7 & 558 & 373 & -0.572710 \\
        $t_{1e}$,$t_{1p}$,$t_{2ee}$,$t_{2ep}$,$t_{3eep}$ & 475 & 366 & 227 & 14 & 1082 & 727 & -0.572714 \\
        \hline
        LUCJ ansatz (numerical) & 55 & 34 & 20 & 3 & 112 & 43 & -0.569178 \\
        LUCJ ansatz (experimental) & & & & & & & $-0.55 \pm 0.03$ \\
        \hline
        NEO-HF (classical) & & & & & & & -0.558727 \\
        NEO-FCI (classical) & & & & & & & -0.572838 \\
        \hline
    \end{tabular}
\end{table}

Intermediate levels of correlation can be recovered by including double excitations. For $\mathrm{HHq}$, using $\{t_{1e}, t_{2ee}\}$ yields an energy of $-1.079406~\mathrm{Ha}$, sufficient for chemical accuracy and closely matching the NEO-FCI limit. Adding mixed electron--proton double excitations $\{t_{2ee}, t_{2ep}\}$ further lowers the energy by approximately $15~\mathrm{\mu Ha}$, reducing the deviation from the FCI value to about $13~\mathrm{\mu Ha}$. This agrees with previous results for $\mathrm{H_2}$ in a minimal STO-3G basis, where inclusion of the double electronic excitation operator alone recovers nearly the full correlation energy~\cite{AdaptVQE2019_original}.  

For $\mathrm{HHq}$, the electron--proton correlation contribution is significantly smaller than the electron--electron correlation term ($10^{-5}$ vs.\ $10^{-2}$~Ha), due to the large proton mass. In contrast, in $\mathrm{PsH}$ the electronic--positronic correlation energy is comparable to the electron--electron correlation energy (i.e., $E_{\{t_{2ee}, t_{2ep}\}} = -0.572710~\mathrm{Ha}$ relative to   $E_{\{t_{1e}, t_{2ee}\}} = -0.569124~\mathrm{Ha}$ is similar to the latter relative to $E_{\{t_{1e}, t_{1p}\}} = -0.558727~\mathrm{Ha}$). This arises from the comparable masses of the electron and positron, which yield matrix elements of similar magnitude. In $\mathrm{HHq}$, the corresponding mixed terms are suppressed by the proton--electron mass ratio ($\sim 2000$), reducing the overall protonic correlation contribution.
These trends suggest that a balanced trade-off between computational cost and accuracy can be achieved by including electronic and mixed electron–positron double excitations, which effectively capture correlation while maintaining feasible circuit depth.

To relate circuit complexity to hardware constraints, we adopt the heuristic introduced by Leymann and Barzen~\cite{leymann_bitter_2020},
\begin{equation}
    d \cdot w \ll \frac{1}{\epsilon},
\end{equation}
where $d$ and $w$ denote circuit depth and width, respectively, and $\epsilon$ is the average gate error rate. This inequality expresses the qualitative requirement for executing a quantum circuit before decoherence dominates. While this metric does not prescribe an accuracy target, the statistical precision of energy measurements can be improved by increasing the number of measurement shots.  

As an illustrative estimate, approximately 170 gates would be required to achieve a target precision of 0.001~Ha for a six-qubit circuit. However, in practice, even single-excitation UCC ansätze already entail $\sim 47$ CNOT gates, exceeding the realistic depth limits of current NISQ hardware. Consequently, accurate beyond-BO quantum simulations remain computationally prohibitive on existing devices. Detailed resource requirements and corresponding energy estimates are provided in Table~\ref{tab:hhq_PsH_JW}.
   
\subsection{Demonstration on IBM Q}

To demonstrate the feasibility of multicomponent quantum simulations on real hardware, we employed the Local Unitary Cluster Jastrow (LUCJ) ansatz~\cite{motta_bridging_2023}, a variational wavefunction specifically designed for correlated electronic ground states on near-term quantum processors. The LUCJ ansatz captures both dynamic and static correlation effects while substantially reducing circuit depth and two-qubit gate requirements compared to traditional quantum chemistry ansätze such as quadratic unitary coupled cluster singles and doubles (qUCCSD).  

Unlike qUCCSD, which involves deep, nonlocal circuits with many variational parameters, the LUCJ ansatz starts from a restricted Hartree--Fock reference and applies a physically motivated correlator inspired by the repulsive Hubbard model. By penalizing double occupancy on the same spatial orbital (opposite-spin electrons) and restricting correlations to local orbital neighborhoods, the LUCJ formulation balances accuracy and hardware efficiency.

\begin{figure}[t]
    \centering
    \includegraphics[width=0.5\linewidth]{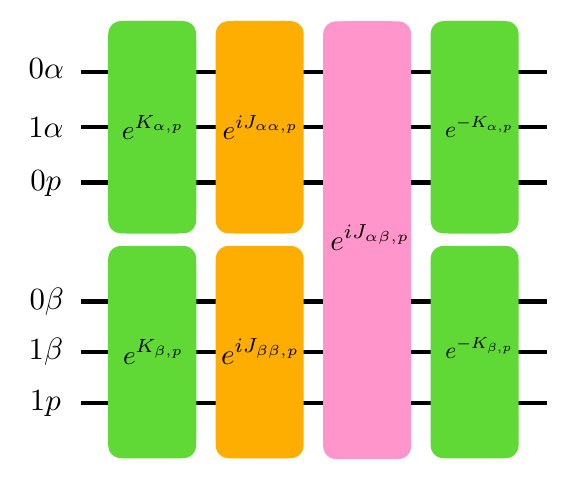}
    \caption{Example circuit block for the LUCJ ansatz applied to $\mathrm{PsH}$, following Eq.~\ref{eq:LUCJansatz}.  
    Orbitals $\{0,1\}$ correspond to the two electronic spatial orbitals, while $p$ labels the positron without any spin orbital. 
    The complete circuit decomposition is shown in Fig.~\ref{fig:topology_circuit}.}
    \label{fig:LUCJ_block}
\end{figure}

\begin{figure}[t]
    \centering
    \includegraphics[width=\linewidth]{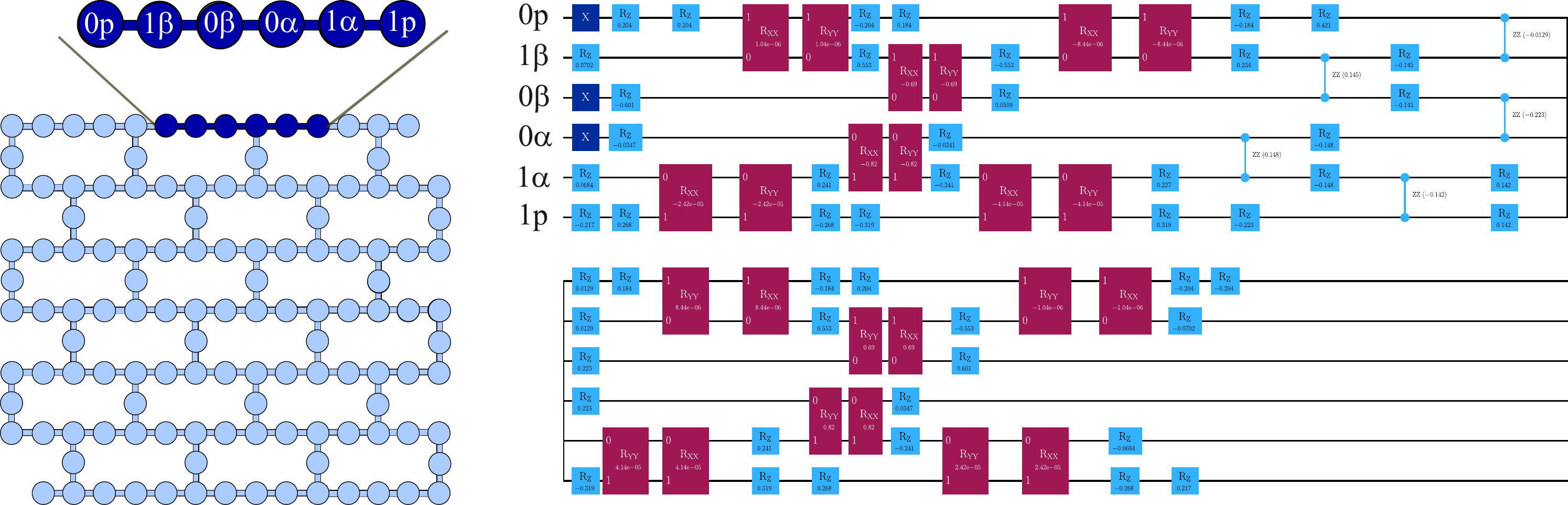}
    \caption{(Left) Topology of the 133-qubit IBM Heron superconducting processor (ibm torino); the 6-qubit subset used for the demonstration is highlighted in dark blue.  
    (Right) LUCJ circuit expressed in the $\{\mathrm{rz},\mathrm{rxx},\mathrm{ryy},\mathrm{rzz},\mathrm{x}\}$ gate basis.  
    Each qubit corresponds to one spatial or spin orbital, and the circuit is initialized in the NEO-HF reference state.}
    \label{fig:topology_circuit}
\end{figure}

The general LUCJ wavefunction is constructed as a product of $L$ local layers:
\begin{equation}
    |\Psi\rangle = \prod_{\mu=1}^{L} e^{\hat{K}_{\mu}} e^{i\hat{J}_{\mu}} e^{-\hat{K}_{\mu}} |\Psi_{\text{NEO-HF}} \rangle,
    \label{eq:LUCJansatz}
\end{equation}
where
\begin{equation}
    \hat{K}_{\mu} = \sum_{s,\sigma} K_{s}^{\mu} \hat{a}_{s\sigma}^{\dagger} \hat{a}_{s\sigma}, 
    \qquad
    \hat{J}_{\mu} = \sum_{s,\sigma\tau} J_{s,\sigma\tau}^{\mu} \hat{n}_{s\sigma} \hat{n}_{s\tau}.
\end{equation}
Here, $\hat{K}_{\mu}$ and $\hat{J}_{\mu}$ represent one-body rotation and two-body number--number correlation operators, respectively, with $s$ indexing spatial orbitals and $\sigma,\tau$ denoting spin.  
For simplicity, all quantum particles (electrons, positrons, and quantum nuclei) are represented by a unified orbital index $s$. This convention is followed in Figs.~\ref{fig:LUCJ_block} and~\ref{fig:topology_circuit}.

Because the circuit depth required by UCC-based approaches remains prohibitive for current hardware, we employ a single LUCJ layer ($L=1$) to approximate the ground-state energies of $\mathrm{HHq}$ and $\mathrm{PsH}$.  
This ansatz achieves a $\sim57\%$ reduction in CNOT gate count relative to the minimal UCC ansatz including $\{t_{1e}, t_{1p}\}$ excitations, while maintaining comparable accuracy (see Table~\ref{tab:hhq_PsH_JW}).

The LUCJ circuits were implemented and executed on the 133-qubit IBM Q Heron superconducting processor (ibm torino).  
A six-qubit subset was selected, as shown in Fig.~\ref{fig:topology_circuit}, to represent the low- and high-energy electron spin orbitals ($0\alpha, 0\beta, 1\alpha, 1\beta$) and the positron/nucleus spatial orbitals (0p and 1p).  
Only local two-qubit operations between adjacent qubits were used, respecting the device topology and minimizing cross-talk.

\subsection{Quantum Error Mitigation via Physics-Inspired Extrapolation (PIE).}
To mitigate hardware noise, we implemented the recently proposed Physics-Inspired Extrapolation (PIE) method~\cite{DiezValle_2025}.  
PIE builds on the Error Mitigation by Restricted Evolution (EMRE) framework~\cite{Saxena2024Sep}, which provides an analytical form for the extrapolation function used to recover noise-free observables.  
Unlike polynomial extrapolation, PIE yields interpretable extrapolation parameters, constant runtime scaling, and reduced sampling overhead.  
In PIE, noise is systematically amplified via circuit folding, and expectation values are measured at multiple noise levels.  
The results are then extrapolated to the zero-noise limit using a linearized model derived from EMRE.  
Depending on circuit complexity, full or partial folding is employed to achieve controlled noise amplification.

\begin{figure}[th]
    \centering
    \includegraphics[width=\linewidth]{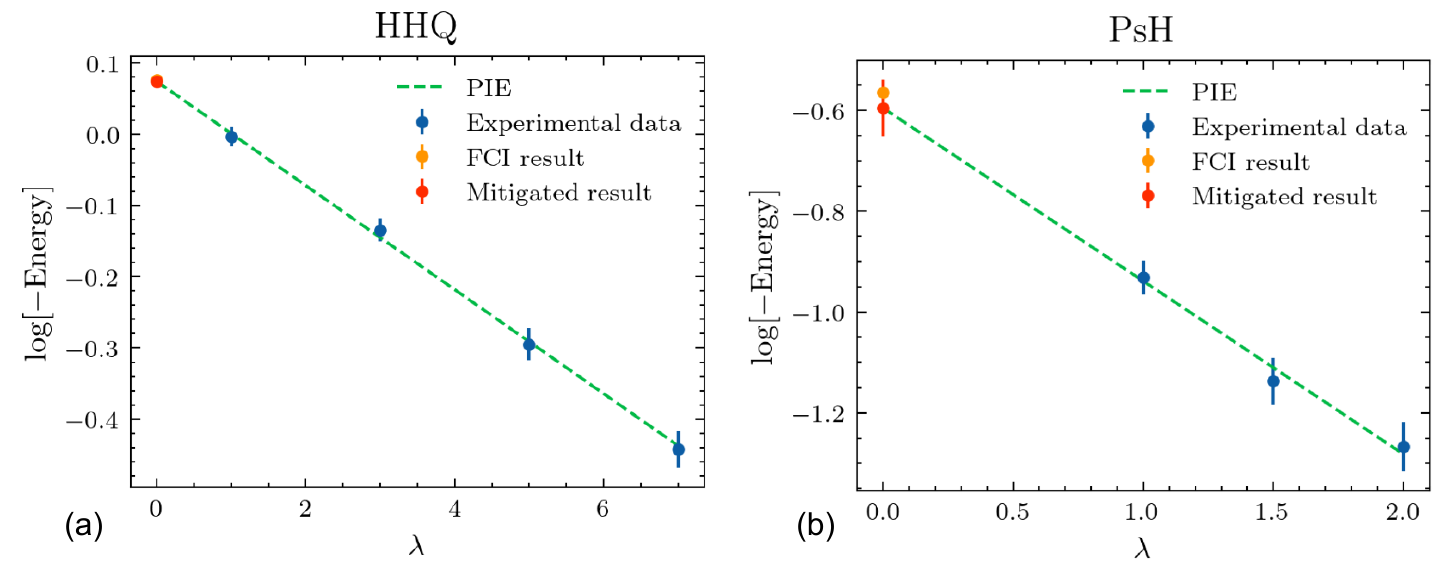}
    \caption{Extrapolated energy results using the PIE method for $\mathrm{HHq}$ and $\mathrm{PsH}$.  
    The logarithm of the negative energy is plotted as a function of the number of circuit foldings, with the zero-noise limit obtained from a linear fit. Raw quantum circuit executions correspond to experimental data at $\lambda=1$, while the noise-mitigated energies are denoted by red dots at $\lambda=0$.}
    \label{fig:IBM_results}
\end{figure}

The experimental execution of the LUCJ circuits and the implementation of the error mitigation protocol are shown in Fig.~\ref{fig:IBM_results}. Optimal LUCJ parameters were first obtained through noiseless classical simulations, yielding energies of $-1.079406$~Ha for $\mathrm{HHq}$ and $-0.569178$~Ha for $\mathrm{PsH}$.  
The corresponding quantum circuits were then executed on the ibm torino processor using 4096 shots per circuit.  
Noisy runs produced raw energies of $-0.996468 \pm 0.013563$~Ha (HHq) and $-0.393893 \pm 0.013127$~Ha (PsH), indicated by the experimental data at $\lambda=1$ in Fig.~\ref{fig:IBM_results}.  
After applying PIE-based error mitigation, the extrapolated energies improved to $-1.076668 \pm 0.009229$~Ha and $-0.551371 \pm 0.031024$~Ha, respectively. These mitigated results are denoted by red dots at $\lambda=0$ in Fig.~\ref{fig:IBM_results}.  
Uncertainties correspond to the standard deviation from repeated runs. The experimental result is significantly lower than the NEO-HF energy for HHq, but it is similar to the NEO-HF energy for PsH, although the uncertainties encompass the NEO-FCI energy.

The error-mitigated results achieved with the LUCJ ansatz and PIE are chemically accurate, closely matching the classical VQE results obtained using the $\{t_{1e}, t_{2ee}\}$ UCC operator set, while requiring substantially fewer gates.  
Although the results remain several millihartrees above the FCI limit for HHq and around 20 millihartrees above the FCI limit for PsH, this demonstration highlights the viability of combining physically motivated ansätze and advanced error mitigation to achieve multicomponent quantum simulations on current NISQ hardware.

\section{Concluding Remarks \label{sec:concluding_remarks}}

In this work, we carried out multicomponent electronic structure calculations for positronium hydride ($\mathrm{PsH}$) and dihydrogen with a quantum mechanical proton ($\mathrm{HHq}$) using quantum computing frameworks.  
Simulations on a quantum computer emulator enabled us to systematically assess the contributions of different excitation operators within the mcUCC ansatz and to quantify the correlation energy recovered under various truncation schemes.  

In addition, we performed experimental demonstrations on IBM’s superconducting quantum hardware using the Local Unitary Cluster Jastrow ansatz in combination with the Physics-Inspired Extrapolation error mitigation technique.  
This hybrid strategy achieved chemically accurate ground-state energies for both $\mathrm{PsH}$ and $\mathrm{HHq}$ while operating within the resource constraints of current noisy intermediate-scale quantum devices.  
The LUCJ ansatz provided a compact, hardware-efficient alternative to traditional coupled-cluster-based ansätze, and PIE successfully mitigated hardware noise to recover high-fidelity energies from noisy measurements.

Together, these results demonstrate a viable path toward scalable, beyond–Born–Oppenheimer quantum simulations that explicitly incorporate nuclear quantum effects.  
By combining physically motivated ansätze, resource-efficient circuit constructions, and advanced error mitigation, this work establishes a foundation for accurate, multicomponent quantum chemistry on near-term quantum processors and offers a roadmap toward chemically relevant quantum simulations in the NISQ era.

\section{Acknowledgements}
This material is based upon work supported by the National Science Foundation for the Center for Quantum Dynamics on Modular Quantum Devices, grant number CHE-2124511.

\bibliography{refs.bib}

@article{Pavosevic2022Jun,
	author = {Pavo{\ifmmode\check{s}\else\v{s}\fi}evi{\ifmmode\acute{c}\else\'{c}\fi}, Fabijan and Hammes-Schiffer, Sharon},
	title = {{Triple electron-electron-proton excitations and second-order approximations in nuclear-electronic orbital coupled cluster methods}},
	journal = {arXiv},
	year = {2022},
	month = jun,
	eprint = {2206.13616},
	doi = {10.1063/5.0106173}
}

@article{Coester1960Jun,
	author = {Coester, F. and K{\ifmmode\ddot{u}\else\"{u}\fi}mmel, H.},
	title = {{Short-range correlations in nuclear wave functions}},
	journal = {Nuclear Physics},
	volume = {17},
	pages = {477--485},
	year = {1960},
	month = jun,
	issn = {0029-5582},
	publisher = {North-Holland},
	doi = {10.1016/0029-5582(60)90140-1}
}

@article{Anand2022Mar,
	author = {Anand, Abhinav and Schleich, Philipp and Alperin-Lea, Sumner and Jensen, Phillip W. K. and Sim, Sukin and D{\ifmmode\acute{\imath}\else\'{\i}\fi}az-Tinoco, Manuel and Kottmann, Jakob S. and Degroote, Matthias and Izmaylov, Artur F. and Aspuru-Guzik, Al{\ifmmode\acute{a}\else\'{a}\fi}n},
	title = {{A quantum computing view on unitary coupled cluster theory}},
	journal = {Chem. Soc. Rev.},
	volume = {51},
	number = {5},
	pages = {1659--1684},
	year = {2022},
	month = mar,
	issn = {0306-0012},
	publisher = {The Royal Society of Chemistry},
	doi = {10.1039/D1CS00932J}
}

@article{Kandala2017Sep,
	author = {Kandala, Abhinav and Mezzacapo, Antonio and Temme, Kristan and Takita, Maika and Brink, Markus and Chow, Jerry M. and Gambetta, Jay M.},
	title = {{Hardware-efficient variational quantum eigensolver for small molecules and quantum magnets}},
	journal = {Nature},
	volume = {549},
	pages = {242--246},
	year = {2017},
	month = sep,
	issn = {1476-4687},
	publisher = {Nature Publishing Group},
	doi = {10.1038/nature23879}
}

@article{Preskill2018Aug,
	author = {Preskill, John},
	title = {{Quantum Computing in the NISQ era and beyond}},
	journal = {Quantum},
	volume = {2},
	pages = {79},
	year = {2018},
	month = aug,
	publisher = {Verein zur F{\ifmmode\ddot{o}\else\"{o}\fi}rderung des Open Access Publizierens in den Quantenwissenschaften},
	eprint = {1801.00862v3},
	doi = {10.22331/q-2018-08-06-79}
}

@article{Bharti2022Feb,
	author = {Bharti, Kishor and Cervera-Lierta, Alba and Kyaw, Thi Ha and Haug, Tobias and Alperin-Lea, Sumner and Anand, Abhinav and Degroote, Matthias and Heimonen, Hermanni and Kottmann, Jakob S. and Menke, Tim and Mok, Wai-Keong and Sim, Sukin and Kwek, Leong-Chuan and Aspuru-Guzik, Al{\ifmmode\acute{a}\else\'{a}\fi}n},
	title = {{Noisy intermediate-scale quantum algorithms}},
	journal = {Rev. Mod. Phys.},
	volume = {94},
	number = {1},
	pages = {015004},
	year = {2022},
	month = feb,
	publisher = {American Physical Society},
	doi = {10.1103/RevModPhys.94.015004}
}

@article{Saxena2024Sep,
	author = {Saxena, Gaurav and Kyaw, Thi Ha},
	title = {{Error Mitigation by Restricted Evolution}},
	journal = {arXiv},
	year = {2024},
	month = sep,
	eprint = {2409.06636},
	doi = {10.48550/arXiv.2409.06636}
}

@article{HammesSchiffer_2021,
  author    = {Hammes{-}Schiffer, Sharon},
  title     = {Nuclear--electronic orbital methods: Foundations and prospects},
  journal   = {J. Chem. Phys.},
  year      = {2021},
  volume    = {155},
  number    = {3},
  pages     = {030901},
  doi       = {10.1063/5.0053576}
}

@article{Li_2017,
  author    = {Li, Ying and Benjamin, Simon C.},
  title     = {Efficient Variational Quantum Simulator Incorporating Active Error Minimization},
  journal   = {Phys. Rev. X},
  year      = {2017},
  volume    = {7},
  number    = {2},
  pages     = {021050},
  doi       = {10.1103/PhysRevX.7.021050}
}

@article{Temme_2017,
  author    = {Temme, Kristan and Bravyi, Sergey and Gambetta, Jay M.},
  title     = {Error Mitigation for Short-Depth Quantum Circuits},
  journal   = {Phys. Rev. Lett.},
  year      = {2017},
  volume    = {119},
  number    = {18},
  pages     = {180509},
  doi       = {10.1103/PhysRevLett.119.180509}
}

@article{BonetMonroig_2018,
  author    = {Bonet-Monroig, Xavier and Sagastizabal, Ruben and Singh, Mali and O'Brien, Teague E.},
  title     = {Low-Cost Error Mitigation by Symmetry Verification},
  journal   = {Phys. Rev. A},
  year      = {2018},
  volume    = {98},
  number    = {6},
  pages     = {062339},
  doi       = {10.1103/PhysRevA.98.062339}
}

@article{Huggins_2021,
  author    = {Huggins, William J. and McArdle, Sam and O'Brien, Teague E. and Lee, Joonho and Rubin, Nicholas C. and Boixo, Sergio and Whaley, K. Birgitta and Babbush, Ryan and McClean, Jarrod R.},
  title     = {Virtual Distillation for Quantum Error Mitigation},
  journal   = {Phys. Rev. X},
  year      = {2021},
  volume    = {11},
  number    = {4},
  pages     = {041036},
  doi       = {10.1103/PhysRevX.11.041036}
}

@article{Arute_2020,
  author    = {Arute, Frank and Arya, Kunal and Babbush, Ryan and others},
  title     = {Hartree--Fock on a Superconducting Qubit Quantum Computer},
  journal   = {Science},
  year      = {2020},
  volume    = {369},
  number    = {6507},
  pages     = {1084--1089},
  doi       = {10.1126/science.abb9811}
}

@article{McCaskey_2019,
  author    = {McCaskey, Alex J. and Parks, Zachary P. and Jakowski, Jacek and Moore, Shirley V. and Morris, Titus D. and Humble, Travis S. and Pooser, Raphael C.},
  title     = {Quantum Chemistry as a Benchmark for Near-Term Quantum Computers},
  journal   = {npj Quantum Information},
  year      = {2019},
  volume    = {5},
  number    = {1},
  pages     = {99},
  doi       = {10.1038/s41534-019-0209-0}
}

@article{Lolur_2023,
  author    = {Lolur, Phalgun and Skogh, M{\aa}rten and Dobrautz, Werner and Warren, Christopher and Bizn{\'a}rov{\'a}, Janka and Osman, Amr and Tancredi, Giovanna and Wendin, G{\"o}ran and Bylander, Jonas and Rahm, Martin},
  title     = {Reference-State Error Mitigation: A Strategy for High Accuracy Quantum Computation of Chemistry},
  journal   = {J. Chem. Theory Comput.},
  year      = {2023},
  volume    = {19},
  number    = {3},
  pages     = {783--789},
  doi       = {10.1021/acs.jctc.2c00807}
}

@misc{DiezValle_2025,
  author    = {D{\'i}ez-Valle, Pablo and Saxena, Gaurav and Baker, Jack S. and Lee, Jun-Ho and Kyaw, Thi Ha},
  title     = {Physics-Inspired Extrapolation for efficient error mitigation and hardware certification},
  howpublished = {arXiv:2505.07977},
  year      = {2025},
  doi = {10.48550/arXiv.2505.07977}
}

@article{mcUCC_NEO,
author = {Pavošević, Fabijan and Hammes-Schiffer, Sharon},
title = {Multicomponent Unitary Coupled Cluster and Equation-of-Motion for Quantum Computation},
journal = {Journal of Chemical Theory and Computation},
volume = {17},
number = {6},
pages = {3252-3258},
year = {2021},
doi = {10.1021/acs.jctc.1c00220},
    note ={PMID: 33945684},

URL = { 
        https://doi.org/10.1021/acs.jctc.1c00220
    
},
eprint = { 
        https://doi.org/10.1021/acs.jctc.1c00220
    
}

}

@article{leymann_bitter_2020,
	title = {The bitter truth about gate-based quantum algorithms in the {NISQ} era},
	volume = {5},
	issn = {2058-9565},
	url = {https://iopscience.iop.org/article/10.1088/2058-9565/abae7d},
	doi = {10.1088/2058-9565/abae7d},
	language = {en},
	number = {4},
	urldate = {2022-12-09},
	journal = {Quantum Science and Technology},
	author = {Leymann, Frank and Barzen, Johanna},
	month = oct,
	year = {2020},
	pages = {044007},
}

@article{bartlett_alternative_1989,
	title = {Alternative {Coupled}-{Cluster} {Ansätze} {II}. {The} {Unitary} {Coupled}-{Cluster} {Method}},
	volume = {155},
	journal = {Chem. Phys. Lett.},
	author = {Bartlett, R. J. and Kucharski, S. A. and Noga, J.},
	year = {1989},
	pages = {133},
}

@article{hempel_quantum_2018,
	title = {Quantum {Chemistry} {Calculations} on a {Trapped}-{Ion} {Quantum} {Simulator}},
	volume = {8},
	journal = {Phys. Rev. X},
	author = {Hempel, C. and Maier, C. and Romero, J. and McClean, J. and Monz, T. and Shen, H. and Jurcevic, P. and Lanyon, B. P. and Love, P. and Babbush, R.},
	year = {2018},
	pages = {031022},
}

@article{romero_strategies_2019,
	title = {Strategies for {Quantum} {Computing} {Molecular} {Energies} {Using} the {Unitary} {Coupled} {Cluster} {Ansatz}},
	volume = {4},
	journal = {Quantum Sci. Technol.},
	author = {Romero, J. and Babbush, R. and McClean, J. R. and Hempel, C. and Love, P. J. and Aspuru-Guzik, A.},
	year = {2019},
	pages = {014008},
}

@article{omalley_scalable_2016,
	title = {Scalable {Quantum} {Simulation} of {Molecular} {Energies}},
	volume = {6},
	journal = {Phys. Rev. X},
	author = {O’Malley, P. J. J. and Babbush, R. and Kivlichan, I. D. and Romero, J. and McClean, J. R. and Barends, R. and Kelly, J. and Roushan, P. and Tranter, A. and Ding, N.},
	year = {2016},
	pages = {031007},
}

@article{peruzzo_variational_2014,
	title = {A {Variational} {Eigenvalue} {Solver} on a {Photonic} {Quantum} {Processor}},
	volume = {5},
	journal = {Nat. Commun.},
	author = {Peruzzo, A. and McClean, J. and Shadbolt, P. and Yung, M.-H. and Zhou, X.-Q. and Love, P. J. and Aspuru-Guzik, A. and O’brien, J. L.},
	year = {2014},
	pages = {4213},
}

@article{cao_quantum_2019,
	title = {Quantum {Chemistry} in the {Age} of {Quantum} {Computing}},
	volume = {119},
	journal = {Chem. Rev.},
	author = {Cao, Y. and Romero, J. and Olson, J. P. and Degroote, M. and Johnson, P. D. and Kieferová, M. and Kivlichan, I. D. and Menke, T. and Peropadre, B. and Sawaya, N. P. D. and Sukin, Sim and Veis, L. and Aspuru-Guzik, A.},
	year = {2019},
	pages = {10856},
}

@article{abrams_simulation_1997_hubbard,
	title = {Simulation of {Many}-{Body} {Fermi} {Systems} on a {Universal} {Quantum} {Computer}},
	volume = {79},
	journal = {Phys. Rev. Lett.},
	author = {Abrams, D. S. and Lloyd, S.},
	year = {1997},
	pages = {2586},
}

@article{aspuru-guzik_simulated_2005,
	title = {Simulated {Quantum} {Computation} of {Molecular} {Energies}},
	volume = {309},
	journal = {Science},
	author = {Aspuru-Guzik, A. and Dutoi, A. D. and Love, P. J. and Head-Gordon, M.},
	year = {2005},
	pages = {1704},
}

@article{bauer_quantum_2020,
	title = {Quantum {Algorithms} for {Quantum} {Chemistry} and {Quantum} {Materials} {Science}},
	volume = {120},
	journal = {Chem. Rev.},
	author = {Bauer, B. and Bravyi, S. and Motta, M. and Kin-Lic Chan, G.},
	year = {2020},
	pages = {12685},
}

@Article{AdaptVQE2019_original,
author={Grimsley, Harper R.
and Economou, Sophia E.
and Barnes, Edwin
and Mayhall, Nicholas J.},
title={An adaptive variational algorithm for exact molecular simulations on a quantum computer},
journal={Nature Communications},
year={2019},
month={Jul},
day={08},
volume={10},
number={1},
pages={3007},
issn={2041-1723},
doi={10.1038/s41467-019-10988-2},
url={https://doi.org/10.1038/s41467-019-10988-2}
}

@misc{ Qiskit,
       author = {MD SAJID ANIS and Abby-Mitchell and H{\'e}ctor Abraham and AduOffei and Rochisha Agarwal and Gabriele Agliardi and Merav Aharoni and Vishnu Ajith and Ismail Yunus Akhalwaya and Gadi Aleksandrowicz and Thomas Alexander and Matthew Amy and Sashwat Anagolum and Anthony-Gandon and Israel F. Araujo and Eli Arbel and Abraham Asfaw and Anish Athalye and Artur Avkhadiev and Carlos Azaustre and PRATHAMESH BHOLE and Abhik Banerjee and Santanu Banerjee and Will Bang and Aman Bansal and Panagiotis Barkoutsos and Ashish Barnawal and George Barron and George S. Barron and Luciano Bello and Yael Ben-Haim and M. Chandler Bennett and Daniel Bevenius and Dhruv Bhatnagar and Prakhar Bhatnagar and Arjun Bhobe and Paolo Bianchini and Lev S. Bishop and Carsten Blank and Sorin Bolos and Soham Bopardikar and Samuel Bosch and Sebastian Brandhofer and Brandon and Sergey Bravyi and Nick Bronn and Bryce-Fuller and David Bucher and Artemiy Burov and Fran Cabrera and Padraic Calpin and Lauren Capelluto and Jorge Carballo and Gin{\'e}s Carrascal and Adam Carriker and Ivan Carvalho and Adrian Chen and Chun-Fu Chen and Edward Chen and Jielun (Chris) Chen and Richard Chen and Franck Chevallier and Kartik Chinda and Rathish Cholarajan and Jerry M. Chow and Spencer Churchill and CisterMoke and Christian Claus and Christian Clauss and Caleb Clothier and Romilly Cocking and Ryan Cocuzzo and Jordan Connor and Filipe Correa and Zachary Crockett and Abigail J. Cross and Andrew W. Cross and Simon Cross and Juan Cruz-Benito and Chris Culver and Antonio D. C{\'o}rcoles-Gonzales and Navaneeth D and Sean Dague and Tareq El Dandachi and Animesh N Dangwal and Jonathan Daniel and Marcus Daniels and Matthieu Dartiailh and Abd{\'o}n Rodr{\'\i}guez Davila and Faisal Debouni and Anton Dekusar and Amol Deshmukh and Mohit Deshpande and Delton Ding and Jun Doi and Eli M. Dow and Patrick Downing and Eric Drechsler and Eugene Dumitrescu and Karel Dumon and Ivan Duran and Kareem EL-Safty and Eric Eastman and Grant Eberle and Amir Ebrahimi and Pieter Eendebak and Daniel Egger and ElePT and Emilio and Alberto Espiricueta and Mark Everitt and Davide Facoetti and Farida and Paco Mart{\'\i}n Fern{\'a}ndez and Samuele Ferracin and Davide Ferrari and Axel Hern{\'a}ndez Ferrera and Romain Fouilland and Albert Frisch and Andreas Fuhrer and Bryce Fuller and MELVIN GEORGE and Julien Gacon and Borja Godoy Gago and Claudio Gambella and Jay M. Gambetta and Adhisha Gammanpila and Luis Garcia and Tanya Garg and Shelly Garion and James R. Garrison and Jim Garrison and Tim Gates and Hristo Georgiev and Leron Gil and Austin Gilliam and Aditya Giridharan and Glen and Juan Gomez-Mosquera and Gonzalo and Salvador de la Puente Gonz{\'a}lez and Jesse Gorzinski and Ian Gould and Donny Greenberg and Dmitry Grinko and Wen Guan and Dani Guijo and Guillermo-Mijares-Vilarino and John A. Gunnels and Harshit Gupta and Naman Gupta and Jakob M. G{\"u}nther and Mikael Haglund and Isabel Haide and Ikko Hamamura and Omar Costa Hamido and Frank Harkins and Kevin Hartman and Areeq Hasan and Vojtech Havlicek and Joe Hellmers and {\L}ukasz Herok and Stefan Hillmich and Colin Hong and Hiroshi Horii and Connor Howington and Shaohan Hu and Wei Hu and Chih-Han Huang and Junye Huang and Rolf Huisman and Haruki Imai and Takashi Imamichi and Kazuaki Ishizaki and Ishwor and Raban Iten and Toshinari Itoko and Alexander Ivrii and Ali Javadi and Ali Javadi-Abhari and Wahaj Javed and Qian Jianhua and Madhav Jivrajani and Kiran Johns and Scott Johnstun and Jonathan-Shoemaker and JosDenmark and JoshDumo and John Judge and Tal Kachmann and Akshay Kale and Naoki Kanazawa and Jessica Kane and Kang-Bae and Annanay Kapila and Anton Karazeev and Paul Kassebaum and Tobias Kehrer and Josh Kelso and Scott Kelso and Hugo van Kemenade and Vismai Khanderao and Spencer King and Yuri Kobayashi and Kovi11Day and Arseny Kovyrshin and Rajiv Krishnakumar and Pradeep Krishnamurthy and Vivek Krishnan and Kevin Krsulich and Prasad Kumkar and Gawel Kus and Ryan LaRose and Enrique Lacal and Rapha{\"e}l Lambert and Haggai Landa and John Lapeyre and Joe Latone and Scott Lawrence and Christina Lee and Gushu Li and Tan Jun Liang and Jake Lishman and Dennis Liu and Peng Liu and Lolcroc and Abhishek K M and Liam Madden and Yunho Maeng and Saurav Maheshkar and Kahan Majmudar and Aleksei Malyshev and Mohamed El Mandouh and Joshua Manela and Manjula and Jakub Marecek and Manoel Marques and Kunal Marwaha and Dmitri Maslov and Pawe{\l} Maszota and Dolph Mathews and Atsushi Matsuo and Farai Mazhandu and Doug McClure and Maureen McElaney and Joseph McElroy and Cameron McGarry and David McKay and Dan McPherson and Srujan Meesala and Dekel Meirom and Corey Mendell and Thomas Metcalfe and Martin Mevissen and Andrew Meyer and Antonio Mezzacapo and Rohit Midha and Daniel Miller and Hannah Miller and Zlatko Minev and Abby Mitchell and Nikolaj Moll and Alejandro Montanez and Gabriel Monteiro and Michael Duane Mooring and Renier Morales and Niall Moran and David Morcuende and Seif Mostafa and Mario Motta and Romain Moyard and Prakash Murali and Daiki Murata and Jan M{\"u}ggenburg and Tristan NEMOZ and David Nadlinger and Ken Nakanishi and Giacomo Nannicini and Paul Nation and Edwin Navarro and Yehuda Naveh and Scott Wyman Neagle and Patrick Neuweiler and Aziz Ngoueya and Thien Nguyen and Johan Nicander and Nick-Singstock and Pradeep Niroula and Hassi Norlen and NuoWenLei and Lee James O'Riordan and Oluwatobi Ogunbayo and Pauline Ollitrault and Tamiya Onodera and Raul Otaolea and Steven Oud and Dan Padilha and Hanhee Paik and Soham Pal and Yuchen Pang and Ashish Panigrahi and Vincent R. Pascuzzi and Simone Perriello and Eric Peterson and Anna Phan and Kuba Pilch and Francesco Piro and Marco Pistoia and Christophe Piveteau and Julia Plewa and Pierre Pocreau and Alejandro Pozas-Kerstjens and Rafa{\l} Pracht and Milos Prokop and Viktor Prutyanov and Sumit Puri and Daniel Puzzuoli and Pythonix and Jes{\'u}s P{\'e}rez and Quant02 and Quintiii and Rafey Iqbal Rahman and Arun Raja and Roshan Rajeev and Isha Rajput and Nipun Ramagiri and Anirudh Rao and Rudy Raymond and Oliver Reardon-Smith and Rafael Mart{\'\i}n-Cuevas Redondo and Max Reuter and Julia Rice and Matt Riedemann and Rietesh and Drew Risinger and Pedro Rivero and Marcello La Rocca and Diego M. Rodr{\'\i}guez and RohithKarur and Ben Rosand and Max Rossmannek and Mingi Ryu and Tharrmashastha SAPV and Nahum Rosa Cruz Sa and Arijit Saha and Abdullah Ash- Saki and Sankalp Sanand and Martin Sandberg and Hirmay Sandesara and Ritvik Sapra and Hayk Sargsyan and Aniruddha Sarkar and Ninad Sathaye and Niko Savola and Bruno Schmitt and Chris Schnabel and Zachary Schoenfeld and Travis L. Scholten and Eddie Schoute and Mark Schulterbrandt and Joachim Schwarm and James Seaward and Sergi and Ismael Faro Sertage and Kanav Setia and Freya Shah and Nathan Shammah and Will Shanks and Rohan Sharma and Polly Shaw and Yunong Shi and Jonathan Shoemaker and Adenilton Silva and Andrea Simonetto and Deeksha Singh and Divyanshu Singh and Parmeet Singh and Phattharaporn Singkanipa and Yukio Siraichi and Siri and Jes{\'u}s Sistos and Iskandar Sitdikov and Seyon Sivarajah and Slavikmew and Magnus Berg Sletfjerding and John A. Smolin and Mathias Soeken and Igor Olegovich Sokolov and Igor Sokolov and Vicente P. Soloviev and SooluThomas and Starfish and Dominik Steenken and Matt Stypulkoski and Adrien Suau and Shaojun Sun and Kevin J. Sung and Makoto Suwama and Oskar S{\l}owik and Rohit Taeja and Hitomi Takahashi and Tanvesh Takawale and Ivano Tavernelli and Charles Taylor and Pete Taylour and Soolu Thomas and Kevin Tian and Mathieu Tillet and Maddy Tod and Miroslav Tomasik and Caroline Tornow and Enrique de la Torre and Juan Luis S{\'a}nchez Toural and Kenso Trabing and Matthew Treinish and Dimitar Trenev and TrishaPe and Felix Truger and Georgios Tsilimigkounakis and Davindra Tulsi and Do{\u{g}}ukan Tuna and Wes Turner and Yotam Vaknin and Carmen Recio Valcarce and Francois Varchon and Adish Vartak and Almudena Carrera Vazquez and Prajjwal Vijaywargiya and Victor Villar and Bhargav Vishnu and Desiree Vogt-Lee and Christophe Vuillot and James Weaver and Johannes Weidenfeller and Rafal Wieczorek and Jonathan A. Wildstrom and Jessica Wilson and Erick Winston and WinterSoldier and Jack J. Woehr and Stefan Woerner and Ryan Woo and Christopher J. Wood and Ryan Wood and Steve Wood and James Wootton and Matt Wright and Lucy Xing and Jintao YU and Yaiza and Bo Yang and Unchun Yang and Jimmy Yao and Daniyar Yeralin and Ryota Yonekura and David Yonge-Mallo and Ryuhei Yoshida and Richard Young and Jessie Yu and Lebin Yu and Yuma-Nakamura and Christopher Zachow and Laura Zdanski and Helena Zhang and Iulia Zidaru and Bastian Zimmermann and Christa Zoufal and aeddins-ibm and alexzhang13 and b63 and bartek-bartlomiej and bcamorrison and brandhsn and chetmurthy and choerst-ibm and deeplokhande and dekel.meirom and dime10 and dlasecki and ehchen and ewinston and fanizzamarco and fs1132429 and gadial and galeinston and georgezhou20 and georgios-ts and gruu and hhorii and hhyap and hykavitha and itoko and jeppevinkel and jessica-angel7 and jezerjojo14 and jliu45 and johannesgreiner and jscott2 and kUmezawa and klinvill and krutik2966 and ma5x and michelle4654 and msuwama and nico-lgrs and nrhawkins and ntgiwsvp and ordmoj and sagar pahwa and pritamsinha2304 and rithikaadiga and ryancocuzzo and saktar-unr and saswati-qiskit and septembrr and sethmerkel and sg495 and shaashwat and smturro2 and sternparky and strickroman and tigerjack and tsura-crisaldo and upsideon and vadebayo49 and welien and willhbang and wmurphy-collabstar and yang.luh and yuri@FreeBSD and Mantas {\v{C}}epulkovskis},
       title = {Qiskit: An Open-source Framework for Quantum Computing},
       year = {2021},
       doi = {10.5281/zenodo.2573505}
}

@article{openfermion,
	doi = {10.1088/2058-9565/ab8ebc},
	url = {https://doi.org/10.1088/2058-9565/ab8ebc},
	year = 2020,
	month = {jun},
	publisher = {{IOP} Publishing},
	volume = {5},
	number = {3},
	pages = {034014},
	author = {Jarrod R McClean and Nicholas C Rubin and Kevin J Sung and Ian D Kivlichan and Xavier Bonet-Monroig and Yudong Cao and Chengyu Dai and E Schuyler Fried and Craig Gidney and Brendan Gimby and Pranav Gokhale and Thomas Häner and Tarini Hardikar and Vojt{\v{e}}ch Havl{\'{\i}}{\v{c}}ek and Oscar Higgott and Cupjin Huang and Josh Izaac and Zhang Jiang and Xinle Liu and Sam McArdle and Matthew Neeley and Thomas O'Brien and Bryan O'Gorman and Isil Ozfidan and Maxwell D Radin and Jhonathan Romero and Nicolas P D Sawaya and Bruno Senjean and Kanav Setia and Sukin Sim and Damian S Steiger and Mark Steudtner and Qiming Sun and Wei Sun and Daochen Wang and Fang Zhang and Ryan Babbush},
	title = {{OpenFermion}: the electronic structure package for quantum computers},
	journal = {Quantum Science and Technology}
}

@article{Bravyi_Kitaev_doi:10.1063/1.4768229,
author = {Seeley,Jacob T.  and Richard,Martin J.  and Love,Peter J. },
title = {The Bravyi-Kitaev transformation for quantum computation of electronic structure},
journal = {The Journal of Chemical Physics},
volume = {137},
number = {22},
pages = {224109},
year = {2012},
doi = {10.1063/1.4768229},

URL = { 
        https://doi.org/10.1063/1.4768229
    
},
eprint = { 
        https://doi.org/10.1063/1.4768229
    
}
}

@article{JordanWigner1928,
author={Jordan, P.
and Wigner, E.},
title={{\"U}ber das Paulische {\"A}quivalenzverbot},
journal={Zeitschrift f{\"u}r Physik},
year={1928},
month={Sep},
day={01},
volume={47},
number={9},
pages={631-651},
issn={0044-3328},
doi={10.1007/BF01331938},
url={https://doi.org/10.1007/BF01331938}
}

@article{JordanWigner_application_PhysRevA.65.042323,
  title = {Simulating physical phenomena by quantum networks},
  author = {Somma, R. and Ortiz, G. and Gubernatis, J. E. and Knill, E. and Laflamme, R.},
  journal = {Phys. Rev. A},
  volume = {65},
  issue = {4},
  pages = {042323},
  numpages = {17},
  year = {2002},
  month = {Apr},
  publisher = {American Physical Society},
  doi = {10.1103/PhysRevA.65.042323},
  url = {https://link.aps.org/doi/10.1103/PhysRevA.65.042323}
}

@article{motta_bridging_2023,
	title = {Bridging physical intuition and hardware efficiency for correlated electronic states: the local unitary cluster Jastrow ansatz for electronic structure},
	volume = {14},
	issn = {2041-6520, 2041-6539},
	url = {https://xlink.rsc.org/?DOI=D3SC02516K},
	doi = {10.1039/D3SC02516K},
	shorttitle = {Bridging physical intuition and hardware efficiency for correlated electronic states},
	pages = {11213--11227},
	number = {40},
	journaltitle = {Chemical Science},
	shortjournal = {Chem. Sci.},
	author = {Motta, Mario and Sung, Kevin J. and Whaley, K. Birgitta and Head-Gordon, Martin and Shee, James},
	urldate = {2025-08-18},
	date = {2023},
	langid = {english},
}

@article{Kovyrshin2023_long,
  title = {A quantum computing implementation of nuclearelectronic orbital (NEO) theory: Toward an exact pre-Born–Oppenheimer formulation of molecular quantum systems},
  volume = {158},
  ISSN = {1089-7690},
  url = {http://dx.doi.org/10.1063/5.0150291},
  DOI = {10.1063/5.0150291},
  number = {21},
  journal = {The Journal of Chemical Physics},
  publisher = {AIP Publishing},
  author = {Kovyrshin,  Arseny and Skogh,  Mårten and Broo,  Anders and Mensa,  Stefano and Sahin,  Emre and Crain,  Jason and Tavernelli,  Ivano},
  year = {2023},
  month = jun 
}

@article{Kovyrshin2023,
  title = {Nonadiabatic Nuclear–Electron Dynamics: A Quantum Computing Approach},
  volume = {14},
  ISSN = {1948-7185},
  url = {http://dx.doi.org/10.1021/acs.jpclett.3c01589},
  DOI = {10.1021/acs.jpclett.3c01589},
  number = {31},
  journal = {The Journal of Physical Chemistry Letters},
  publisher = {American Chemical Society (ACS)},
  author = {Kovyrshin,  Arseny and Skogh,  Mårten and Tornberg,  Lars and Broo,  Anders and Mensa,  Stefano and Sahin,  Emre and Symons,  Benjamin C. B. and Crain,  Jason and Tavernelli,  Ivano},
  year = {2023},
  month = aug,
  pages = {7065–7072}
}

@misc{kovyrshin2025approximatequantumcircuitcompilation,
      title={Approximate quantum circuit compilation for proton-transfer kinetics on quantum processors}, 
      author={Arseny Kovyrshin and Dilhan Manawadu and Edoardo Altamura and George Pennington and Benjamin Jaderberg and Sebastian Brandhofer and Anton Nykänen and Aaron Miller and Walter Talarico and Stefan Knecht and Fabijan Pavošević and Alberto Baiardi and Francesco Tacchino and Ivano Tavernelli and Stefano Mensa and Jason Crain and Lars Tornberg and Anders Broo},
      year={2025},
      eprint={2507.08996},
      archivePrefix={arXiv},
      primaryClass={quant-ph},
      url={https://arxiv.org/abs/2507.08996}, 
}

@article{nykanen2023toward,
  title={Toward accurate post-born--oppenheimer molecular simulations on quantum computers: An adaptive variational eigensolver with nuclear-electronic frozen natural orbitals},
  author={Nykanen, Anton and Miller, Aaron and Talarico, Walter and Knecht, Stefan and Kovyrshin, Arseny and Skogh, Marten and Tornberg, Lars and Broo, Anders and Mensa, Stefano and Symons, Benjamin CB and others},
  journal={Journal of Chemical Theory and Computation},
  volume={19},
  number={24},
  pages={9269--9277},
  year={2023},
  publisher={ACS Publications}
}

@article{Culpitt2025,
  title = {Constrained Nuclear-Electronic Orbital Theory for Quantum Computation},
  volume = {21},
  ISSN = {1549-9626},
  url = {http://dx.doi.org/10.1021/acs.jctc.5c00815},
  DOI = {10.1021/acs.jctc.5c00815},
  number = {16},
  journal = {Journal of Chemical Theory and Computation},
  publisher = {American Chemical Society (ACS)},
  author = {Culpitt,  Tanner and Chen,  Zehua and Pavošević,  Fabijan and Yang,  Yang},
  year = {2025},
  month = aug,
  pages = {7845–7854}
}

@article{Waluk2024,
  title = {Nuclear Quantum Effects in Proton or Hydrogen Transfer},
  volume = {15},
  ISSN = {1948-7185},
  url = {http://dx.doi.org/10.1021/acs.jpclett.3c03368},
  DOI = {10.1021/acs.jpclett.3c03368},
  number = {2},
  journal = {The Journal of Physical Chemistry Letters},
  publisher = {American Chemical Society (ACS)},
  author = {Waluk,  Jacek},
  year = {2024},
  month = jan,
  pages = {598–607}
}

@article{Meisner2016,
  author    = {Meisner, Jan and K{\"a}stner, Johannes},
  title     = {Atom Tunneling in Chemistry},
  journal   = {Angew. Chem. Int. Ed.},
  year      = {2016},
  volume    = {55},
  number    = {18},
  pages     = {5400--5413},
  doi       = {10.1002/anie.201511028}
}

@article{HammesSchiffer2015,
  author    = {Hammes-Schiffer, Sharon},
  title     = {Proton-Coupled Electron Transfer: Moving Together and Charging Forward},
  journal   = {J. Am. Chem. Soc.},
  year      = {2015},
  volume    = {137},
  number    = {28},
  pages     = {8860--8871},
  doi       = {10.1021/jacs.5b04087}
}

@article{Tyburski2021,
  author    = {Tyburski, Robin and Liu, Tianfei and Glover, Starla D. and Hammarstr{\"o}m, Leif},
  title     = {Proton-Coupled Electron Transfer Guidelines, Fair and Square},
  journal   = {J. Am. Chem. Soc.},
  year      = {2021},
  volume    = {143},
  number    = {2},
  pages     = {560--576},
  doi       = {10.1021/jacs.0c09106}
}

@article{Klinman2013,
  author    = {Klinman, Judith P. and Kohen, Amnon},
  title     = {Hydrogen Tunneling Links Protein Dynamics to Enzyme Catalysis},
  journal   = {Annu. Rev. Biochem.},
  year      = {2013},
  volume    = {82},
  pages     = {471--496},
  doi       = {10.1146/annurev-biochem-051710-133623}
}

@article{HammesSchiffer2023,
  author    = {Hammes-Schiffer, Sharon},
  title     = {Exploring Proton-Coupled Electron Transfer at Multiple Scales},
  journal   = {Nat. Comput. Sci.},
  year      = {2023},
  volume    = {3},
  number    = {4},
  pages     = {291--300},
  doi       = {10.1038/s43588-023-00422-5}
}

@article{webb2002multiconfigurational,
  title={Multiconfigurational nuclear-electronic orbital approach: Incorporation of nuclear quantum effects in electronic structure calculations},
  author={Webb, Simon P and Iordanov, Tzvetelin and Hammes-Schiffer, Sharon},
  journal={The Journal of Chemical Physics},
  volume={117},
  number={9},
  pages={4106--4118},
  year={2002},
  publisher={American Institute of Physics}
}

@article{sto3g,
    author = {Hehre, W. J. and Stewart, R. F. and Pople, J. A.},
    title = {Self‐Consistent Molecular‐Orbital Methods. I. Use of Gaussian Expansions of Slater‐Type Atomic Orbitals},
    journal = {The Journal of Chemical Physics},
    volume = {51},
    number = {6},
    pages = {2657-2664},
    year = {1969},
    month = {09},
    issn = {0021-9606},
    doi = {10.1063/1.1672392},
    url = {https://doi.org/10.1063/1.1672392},
    eprint = {https://pubs.aip.org/aip/jcp/article-pdf/51/6/2657/18864378/2657_1_online.pdf},
}

@article{6-31G,
    author = {Ditchfield, R. and Hehre, W. J. and Pople, J. A.},
    title = {Self‐Consistent Molecular‐Orbital Methods. IX. An Extended Gaussian‐Type Basis for Molecular‐Orbital Studies of Organic Molecules},
    journal = {The Journal of Chemical Physics},
    volume = {54},
    number = {2},
    pages = {724-728},
    year = {1971},
    month = {01},
    issn = {0021-9606},
    doi = {10.1063/1.1674902},
    url = {https://doi.org/10.1063/1.1674902},
    eprint = {https://pubs.aip.org/aip/jcp/article-pdf/54/2/724/18872207/724_1_online.pdf},
}

\end{document}